\documentclass[showpacs,showkeys,preprintnumbers,amsmath,amssymb,aps]{revtex4}
\usepackage[dvips]{graphicx}
\begin{document}

\title{A COMPLEX SHELL MODEL REPRESENTATION INCLUDING ANTIBOUND STATES}

\author  {R. Id Betan}
\affiliation{
Departamento de Fisica, FCEIA, UNR,
Avenida Pellegrini 250, 2000 Rosario, Argentina}
\author{R. J. Liotta}
\affiliation{
KTH, Alba Nova University Center, SE-10691 Stockholm, Sweden}
\author{N. Sandulescu}
\affiliation{
Institute of Physics and Nuclear Engineering, 
P.O.Box MG-6, Bucharest-Magurele, Romania}
\author{T. Vertse}
\affiliation{
Institute of Nuclear Research of the Hungarian  Academy of Sciences,
H-4001 Debrecen, Pf. 51, Hungary,\\
University of Debrecen, Faculty of Information Science, H-4010 Debrecen, Pf. 12,
Hungary}
\author{R. Wyss}
\affiliation{
.KTH, Alba Nova University Center, SE-10691 Stockholm, Sweden}

\date{\today}

\begin{abstract}
A generalization of the
Complex Shell Model formalism is presented which includes antibound states
in the basis. These states, together with bound states, Gamow states,
and the continuum background, represented by properly chosen scattering
waves, form a representation where all states are treated on the same footing.
Two-particle states are evaluated within this formalism and observable
two-particle resonances are defined. The formalism is illustrated in the
well known case of $^{11}$Li in its bound ground state and 
in $^{70}$Ca(gs), which is also bound. Both cases are found to have 
a halo structure. These halo structures are described within the generalized 
Complex Shell Model. We investigated the formation of two-particle 
resonances in these nuclei, but no evidence of such resonances was found.
\end{abstract}

\pacs{21.10.Tg, 23.50.+z, 24.10.Eq}

\keywords{Continuum, Resonances, Weakly-bound nuclei, Halos, 
Antibound states}

\maketitle

\section{Introduction}
\label{sec:intro}
One of the most intriguing developments in nuclear physics
is the disclosure of antibound states as important building
blocks to induce the appearance of halos in exotic nuclei.
The explanation of halos as well as the description
of the behaviour of exotic nuclei, which live a short
time and therefore their dynamics is governed by process
occuring in the continuum part of the spectrum,
requires the introduction of
new and powerful models to take into account the complicated
interplay among different degrees of freedom that produce
weakly bound and resonant states.
Such models appeared as soon as the experimental evidence
in exotic nuclei provided a variety of features 
which could not be explained by standard approaches like the shell model
using harmonic oscillator representations.

The lively activity that characterizes this field can be
attested by the rapid development that is taken
place.  It would be outside the scope of this paper to assess 
those models. One of the first reviews is still
rather recent \cite{zhu}. Since then much have been published.
In light nuclei a concise but clear account is in
Ref. \cite{gar}. A more comprehensive treatment (including 
abundant references) can be found in Ref. \cite{suz}.
The Continuum Shell Model is also been applied for this purpose
\cite{oko}.

Among the attemps to describe the structure of rare nuclei 
the Complex Shell Model method (CXSM) \cite{rsm,mic,cxsm}
presents the advantage of incorporating on an equal footing
bound single-particle states as well as resonances and the 
non-resonant continuum.
Recently, in a short publication \cite{ant}, even the elusive 
antibound states have been included in the CXSM basis. 
The purpose of this paper is to present in a more
extensive fashion the work of Ref. \cite{ant}. At the same
time the influence of the different ingredients that determine
the importance of the antibound state will be appraised
by analyzing realistic situations.
The formalism is in Section \ref{sec:form}, applications
are in Section \ref{sec:appl} and a summary and conclusions
are in Section \ref{sec:sum}.

\section{Formalism}
\label{sec:form}
Although the formalism to be used here was given before,
we will present it again with some detail for clarity
of presentation and also because we will deal with 
unfamiliar quantities like antibound states and 
complex probabilities which we would like to clarify 
from the outset. In addition we would like to
display in the presentation the advantages of working in the
complex energy plane, with its implicit loose of
familiar quantum mechanical concepts, as compared with
standard representations on the real energy axis.

With this in mind we start by pointing out that the 
study of processes taking part in the continuum part
of the spectrum may require, by the very nature of the problem,
a time dependent formalism. Therefore the 
quantum description of
the system may become a very hard undertaking. In fact
it may even become an impossible task, since the dynamics of
the problem may be very sensitive to the initial conditions
and the system can easily precipitate into a chaotic regime.
But there are exceptions to this setting.
Thus, the spectrum corresponding to a system of free particles 
is a monotonous and continuum function of the energy as determined 
by the kinetic motion of the particles, and the time-independent
Schr\"odinger equation is suitable to explain this system.

A not so clear situation where time-independent formalisms
can be applied occurs even if the particles 
start to interact with each other under the influence 
of a central field. Here the continuum may acquire
features which depart from the monotony of the kinetic
energy spectrum. The physical meaning of these structures is 
that due to the interactions the system remain
in a certain configuration during a time, i. e. within an
energy interval. In other words, the system is trapped by 
a barrier erected by the interactions as well as by the
centrifugal motion. The structures appearing on the continuum
background are the resonances.
If the barrier is high enough the
system will remain during a long time in a localized
region of space and the dynamics of the process can be
studied within stationary formalisms.   
The question that one may ask is what is meant by "long time"
or "barrier high enough". This question is irrelevant 
in radioactive decay, since measurable mean lives correspond
to very narrow resonances. Thus, for the shortest measurable
radioactive decay (i. e. the widest measurable resonance)
the mean life is at present
$T\approx 10^{-12}$ sec and the width is, according 
to the uncertainty relation, $\Gamma=6.6\times 10^{-10}$ MeV.
Therefore in this time-energy scale the nucleus lives a long
time before decaying and one may assume that the process is
stationary. But this is not the case in all processes
occurrying in the continuum. In particular the formation
of halos could proceed through wide resonances where even
the proper continuum plays a role, as we will see in the
next Section.
On the other hand, if the resonance is very wide the halflive
is very short, indicating that the system is not trapped in a
barrier during a long time
and the process can not be considered stationary.
One can still try to solve the time-independent Schr\"odinger
equation in this situation to gain insight into the limitations
of the problem. Since the system is not trapped by a barrier 
high enough all or parts of its components will soon depart
from the rest. To study this situation in a many-body case
is a difficult task. We therefore start from the simplest
case, that is a particle moving in a central field,
as Gamow did in the beginning of quantum mechanics \cite{Gam28}.

\subsection{The Berggren representation}
\label{sec:begr}

We thus solve the time-independent Schr\"odinger equation imposing
to the wavefunction $w_{n}(r,k_n)$ regularity at origin and
outgoing boundary conditions (the particle departs from
the origin), i. e. we require \cite{Pei38}
\begin{equation}\label{eq:bcond}
\lim_{r\to 0} w_{n}(r,k_n)=0,~~~~
\lim_{r\to \infty} w_{n}(r,k_n)=N_{n}e^{ik_{n}r}
\end{equation}
where $k_n$ is the asymptotic momentum of the state with energy 
eigenvalue $E_n$, i. e.
\begin{equation}\label{eq:cxen}
E_n=\frac{\hbar^2}{2\mu}k^2_{n}
\end{equation}

The wavefunction $w_{n}(r,k_n)$ thus defined can 
be considered a generalization of the 
definition of eigenvectors of the
single  particle  Schr\"odinger equation.
The eigenvalues $E_n$ can now be complex. 
Writing
\begin{equation}\label{eq:cxkn}
k_{n}=\kappa_{n}+i\gamma_{n}
\end{equation}
the eigenvectors belonging to those eigenvalues 
can be classified in four classes, namely:
(a) bound states, for which $\kappa_{n}$=0 and $\gamma_{n}>0$; (b) 
antibound states with $\kappa_{n}$=0, $\gamma_{n}<0$; (c) decay 
resonant states with $\kappa_{n}>0$, $\gamma_{n}<0$ and (d) 
capture resonant states with $\kappa_{n}<0$, $\gamma_{n}<0$. From 
Eq. (\ref{eq:bcond}) one sees that only the bound state wave 
functions do not diverge. 

The imaginary part of the
energy of the states (c) was interpreted by Gamow as 
minus twice the width of the resonance \cite{Gam28} and
therefore these states are usually called Gamow resonances.
However, we will make a distinction between real resonances, having
physical meaning, and other "resonances" which are generally wide and 
therefore do not correspond to any particular observable state.
To avoid having to distinguish between these different
situations any time we refer to the four classes of outgoing states described 
above we will refer to them in general as "poles", since they 
are the poles of the Green function 
\cite{n66} and, therefore, of the S-matrix. 

With the standard definition of scalar product
only the bound states can be normalized in an infinite interval. 
Therefore this definition has to be generalized in order to be able 
to use the generalized "eigenvectors". 
This can only be done if one 
uses a  bi-orthogonal basis. 
As we will see, a consequence of this generalization is that 
the scalar product between two functions is not the integral of one of
the functions times the complex conjugate of the other but rather  
times the function itself, and one has to apply some regularization 
method for calculating the resulting integrals. We will perform this
task by using the complex rotation method \cite{Gya71}.

Berggren found that some of these complex eigenvectors (bound states and
decaying resonances) can be used to
express the Dirac $\delta$-function \cite{b68}.
We will show the main points
of his derivation since we will come back to it frequently.

One can write the Dirac $\delta$-function on the real energy
axis, i. e. within a quantum mechanical framework,  as \cite{n66}
\begin{equation}\label{eq:deln}
\delta(r-r^\prime)=\sum_n w_n(r)w_n(r^{'}) +
\int_0^\infty dE u(r,E)u(r^{'},E)
\end{equation}
where $w_n(r)$ are the bound states wavefunctions and
$u(r,E)$ are scattering states. The integration contour is along 
the real energy axis.
Notice that it appears the wavefunction times itself, and not times
its complex conjugate, although (\ref{eq:deln}) is a distribution to be
applied on the Hilbert space. This is because for 
both bound states and scattering states on the real energy axis
one can choose the phases such that the wavefunctions are real.

\begin{figure}
\begin{center}
\includegraphics[width=0.5\textwidth]{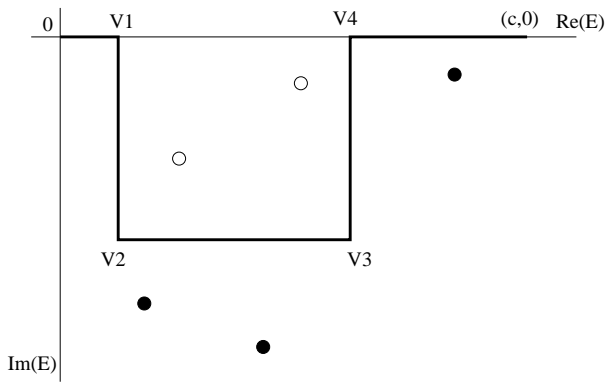}
\vspace{1cm} \caption
{
Integration contour $L^+$ in the complex energy plane as defined
by the vertex points $V_i$. The open circles
represent the Gamow resonances to be included in the sum of Eq. 
(\ref{eq:delb}) while the solid circles are those which are excluded.
The vertex  $(c,0)$ corresponds to the energy cut-off point $c$. 
}
\label{spath}
\end{center}
\end{figure}

Berggren extended the expression (\ref{eq:deln}) by prolonging
the integration contour to the complex energy plane. Using 
the Cauchy theorem one gets \cite{b68} 
\begin{equation}\label{eq:delb}
\delta(r-r^\prime)=\sum_n{\tilde w^*_n(r)}w_n(r^{'}) +
\int_{L^+} dE {\tilde u^*(r,E)}u(r^{'},E)
\end{equation}
where the sum runs over all the bound states plus the complex poles
which lie between the real energy axis and the integration contour 
$L^+$, as shown in Fig. \ref{spath}. This contour may 
have any form one wishes but, since it is a topological deformation 
of the real energy axis, it should be a continuum curve that starts
in the origin, i. e. at $(0,0)$, and end at infinite, i. e. at 
$(\infty,0)$. However, as in any shell model calculation one cuts the
basis at certain maximum energy which in the Figure is the point $(c,0)$.  

The wavefunction  $\tilde w_n(r)$ is the mirror state of
$w_n(r,k_n)$, i. e. the solution with $\tilde k_n = -k^*_n$.
Therefore $\tilde w^*_n(r)=w_n(r)$. The same is valid
for the scattering state  $u(r,E)$. We therefore indeed find
that the internal product is the wavefunction times itself 
and not times its complex conjugate. This internal product is called
Berggren metric.

The Gamow states enclosed by the contour $L^+$, plus the bound states
and the scattering states on the contour, have been shown to form
a complete set of single-particle states (Berggren representation)
to describe many-body states in the complex energy plane \cite{l96}. 

Discretizing the integral in Eq. (\ref{eq:delb}) one obtains the
set of orthonormal vectors $\vert \varphi_j\rangle$ 
forming the Berggren representation. These vectors include
the set of bound and Gamow states, i . e.
$\varphi_p(r)=
\langle r \vert \varphi_p\rangle=\{w_p(r)\}$ and the discretized
scattering states, i. e.
$\varphi_p(r)=
\langle r \vert \varphi_p\rangle=\sqrt{h_p}u(r,E_p)$. The
quantities $E_p$ and $h_p$ are defined by the procedure one uses
to perform the integration. In the Gaussian method $E_p$
are the Gaussian points and
$h_p$ the corresponding weights. 

\subsection{Berggren space and resonances}
\label{sec:bergs}

The representation above spans a space which is called Berggren space.
Since within the metric defining the Berggren space
the definition of scalar product does
not include absolute values, one may have probabilities which 
are complex numbers. We thus find that 
forcing the time-dependent process of particles
interacting in the continuum to be stationary one has to pay the
price of having complex energies and complex probabilities.  
This is not as weird as it may sound, since we are now dealing 
with states lying in the complex energy plane which, in principle,
do not have any physical meaning. This is a point which has
produced some confusion already from the time of the first 
application of the theory nearly twenty years ago \cite{v87}. 
It is therefore important to clarify the meaning of 
this feature here where even another weird feature, namely 
antibound states, will be introduced.

Let us start by pointing out that the Berggren transformation
leading to the representation (\ref{eq:delb}) does not change
in any way the meaning of the Dirac $\delta$-function. 
In particular, since on the real energy axis 
the wavefunctions can be chosen to be real, the results of 
a continuum shell model many-body calculation 
of quantities on the real energy axis (like e. g. sum rules or the energies
of bound states) should coincide with the results provided by the same 
calculation using the Berggren metric with whatever contour one chooses
provided that the resonances enclosed by that contour are also included.
This is a very important property that will be 
used by us to check our results as well as our computer codes.
Only for states lying in the complex energy plane the 
evaluated quantities
will be unfamiliar, which is not surprising since these
states lie outside the Hilbert space. Then the question is why one
uses the Berggren space or, equivalently, which is the advantage of using
the complex energy plane. This is a valid question which we will
therefore answer in some detail.

In absence of any background the usual form of the S-matrix 
corresponding to a partial wave $(lj)$ 
in the neighborhood of an isolated resonance $n$ is
\begin{equation}
\label{eq:bwiso}
{\cal S}_{nlj}(E)=
\frac{E-E_{nlj}^{(0)}-\frac{i}{2}\Gamma_{nlj}}
{E-E_{nlj}^{(0)}+\frac{i}{2}\Gamma_{nlj}}
=\left[1-\frac{i\Gamma_{nlj}}
{E-E_{nlj}^{(0)}+\frac{i}{2}\Gamma_{nlj}}\right]~.
\end{equation}
where $E_{nlj}^{(0)}$ is the position and $\Gamma_{nlj}$ 
the width of the resonance. These are real positive numbers.
 
It is important to point out that Eq. (\ref{eq:bwiso}) is valid
only if the resonance is isolated. This condition is fulfilled for
narrow resonances. In this case 
the residues ${\cal R}$ of the S-matrix is a pure imaginary number,
i. e. ${\cal R}_{nlj}$=-i$\Gamma_{nlj}$, as readily follows 
from Eq. (\ref{eq:bwiso}).

The cross section corresponding to the scattering of a 
particle at an energy close to the isolated resonance 
acquires the form
\begin{equation}
\label{sect}
\sigma_{lj}(E)=(2l+1)\frac{\pi}{k^2}\frac{\Gamma^2_{lj}}
{(E-E^{(0)}_{lj})^2+(\Gamma_{lj}/2)^2}~,
\end{equation}
This formula was derived by G. Breit and E. P. Wigner \cite{Bre36}
to explain the capture of slow neutrons. 
It is one of the most successful expressions ever written in
quantum physics, as shown by its extensive use in the study of 
resonances ever since. It was by comparing with experiment that
Wigner interpreted the number $\Gamma$ as the width of the resonance. 
Since the imaginary part of the S-matrix pole is -$\Gamma/2$
(Eq. (\ref{eq:bwiso})),
this interpretation coincided with the Gamow interpretation of the width.
It is often assumed that the value of $\Gamma$ thus 
defined is the width of the resonance if it is 
narrow, i. e. if $\Gamma$ 
is small enough. However, a state having a very small (in absolute value) 
part of the energy 
is not necessarily a physically meaningful resonance. For instance in the
CXSM calculation of two-particle resonances there are states lying very
close to the real energy axis which form part of the continuum background 
since they are induced by basis vectors belonging to 
the continuum contour \cite{cxsm}. 
But besides this objection one my ask what it is meant by "large" and "small"
width even in the case of a pure Gamow pole.

To avoid these objections and to get a more precise definition of a 
resonance we notice that the condition that the 
resonance is isolated is equivalent to the requirement that the
residues of the S-matrix should be a pure imaginary number. This criterion
was used in Ref. \cite{Ver95} to evaluate partial decay widths
corresponding to the emission of neutrons from giant resonances.
It was thus found that only in a few cases the residues of the S-matrix 
was a pure imaginary number. Usually that quantity was complex and
therefore devoided of any physical meaning. Yet, in some circunstances
one can show that the imaginary part of the width thus evaluated 
(that is, as $\Gamma_{nlj}$=2i${\cal R}_{nlj}$) indicates that the
lifetime of the resonance is too short and that it can be considered
as a part of the background \cite{b78}.

The important conclusion of this discussion is that the imaginary part 
of the energy, which in some cases can even be evaluated analytically 
\cite{Hum61}, is not in itself related to the width. 

A more precise definition of a resonance can be obtained by 
requiring that the corresponding complex pole posseses some 
physical attribute.
As already pointed out, a measurable resonance corresponds to a process
in which the system is trapped inside a barrier during a long time.
One can therefore define a resonance according to its degree of 
localization inside the nuclear volume. This criterion will be 
very important to identify the two-particle resonances to be studied in
the applications. Thus, we will evaluate all resonances that can be
built within our Berggren single-particle representation 
and give physical meaning to 
the ones with wavefunctions showing localization properties within the 
nuclear volume. But it is important to point out that although 
this criterion is more accurate to define a resonance in comparisson
to e. g. the one based on the value of the imaginary part of the 
energy, the very nature of the problem hinders an exact formulation 
of a criterion to define a resonance in the framework of a time 
independent formalism. Therefore even our criterion of defining a 
resonance according to its localization features has to be considered
approximate.

The developing of a resonance depends upon the 
central mean field as well as the two body residual interaction acting
upon the basis elements. As in any shell-model calculation the interplay 
among these elements induces the correlated states which, in our case,
include resonances, bound and antibound states. One thus expects that 
correlated
narrow resonances may be induced not only by bound states and 
narrow resonances but also by wide resonances, antibound states and even the 
continuum itself.
 The evolution of this complicated process as a 
function of the interactions as well as the dimension of the basis 
(including the energy cut-off that defines the representation) can 
clearly be seen in the complex energy plane, since the location of the 
complex energies corresponding to the resonances are just points
in the two-particle complex plane. This would be very difficult to do on 
the real energy axis, since here the wavefunctions corresponding to
wide resonances cannot be easily diferentiated from those
corresponding to the continuum background. This is an important advantage
of using the complex energy plane. We will come back to this point below. 

The two-particle shell-model equations have the standard form, except
the metric, i. e.
\begin{equation}\label{eq:tda}
(\omega_\alpha-\epsilon_i-\epsilon_j)X(ij;\alpha)=\sum_{k\leq l}
 <\tilde{kl};\alpha|V|ij;\alpha> X(kl;\alpha)
\end{equation}
where $\alpha$ labels two-particle states and $i, j, k, l$
label single-particle states. As in Eq. (\ref{eq:delb}) tilde denotes 
mirror states.

In principle the zeroth order energies $\epsilon_i+\epsilon_j$ would 
cover the whole two-particle energy plane (since $\epsilon$ is actually
a continuous variable) and the correlated state would thus be immersed 
in a background of uncorrelated states. However, one can avoid 
this problem by chosen suitable contours \cite{cxsm}.

A convenient fashion of solving Eq. (\ref{eq:tda}) is by using
a separable two-body interaction. One thus obtains the dispersion 
relation \cite{cxsm}
\begin{equation}\label{eq:disrel}
-\frac{1}{G_\alpha} =  \sum_{i\leq j} \frac{f^2_\alpha(ij)}
{\omega_\alpha-\epsilon_i-\epsilon_j}
\end{equation}
where $G_\alpha$ is the interaction strength and  $f_\alpha(ij)$
is the matrix element of the field defining the interaction.
We will choose for this field the derivative of the Woods-Saxon
potential given by 
\begin{equation}\label{eq:wsfield}
F(r)=\frac{f_0}{1+exp(r-R')/a'} 
\end{equation}
i. e. the field $f$ is
\begin{equation}\label{eq:field}
f(r)=-r\frac{dF(r)}{dr} 
\end{equation}
This choice of the field defining the two-body interaction differs 
from the one given in Refs. \cite{rsm,cxsm}, where the function $F$
was the Woods-Saxon potential used to evaluate the single-particle states.
The reason of this is that now the central field defining the 
single-particle states contains also a Gaussian part, as we will see in 
the Aplications. With the present choice of the effective two-body
interaction we are able to describe well experimental data, which is
the main criterion used in Shell Model calculations to define the effective
force. 

The two-particle wavefunction can be written as
\begin{equation}\label{eq:wfa}
 X(ij;\alpha) =  N_\alpha \frac{f_\alpha(ij)}
{\omega_\alpha-\epsilon_i-\epsilon_j}
\end{equation}
where $N_\alpha$ is the normalization constant. 

This form of the wavefunction shows clearly the problem one faces if no 
measure is taken to avoid the continuum background of uncorrelated
states. As one chooses more points on the contour (thus improving, in
principle, the procedure) the energy denominators corresponding
to zeroth order states close to the energy $\omega_\alpha$ diminish
and all wavefunctions tend to a common value, thus making impossible 
the identification of the correlated resonance. 

This possibility of identifying individually the correlated states is another 
favorable feature of the complex energy plane. On the positive side
of the real energy axis (i. e. on the continuum) there is no any unique 
correlated state because the 
poles of the two-particle Green function consist only of bound states,
which lie on the physical $k$-sheet, or to resonances located 
on non-physical sheets \cite{kuk}. Therefore one does not evaluate the 
energy (i. e. the position) of the resonances on the real energy axis 
but rather matrix elements of physical operators, like e. g. transition 
amplitudes, which increase (in absolute value)
close to the resonance energy if the resonance is narrow enough 
\cite{be,ebh}. Such limitations do not exist in the complex 
energy plane. Instead, here one calculates all complex states 
but at the end,
in order to assign physical meaning to those states as well as
to compare with experiment, one chooses the integration contour as
the real energy axis. At this point of the calculation the 
CXSM and standard methods like the Continuum Shell Model coincide.
Only complex states showing the resonant features mentioned above
(at energies around the real part of the complex energy)
will have physical meaning. This is a manifestation that these states
are localized inside the nuclear volume. The wavefunctions of non-localized 
states are small inside the nuclear volumen and their contribution to the 
matrix elements of physical operators is also small \cite{Bia01}. This
feature will be illustrated below, where we will show localized as well
as non-localized states.

Usually one chooses the contour of integration $L^+$ such
that it encompasses only Gamow resonances, as in 
Fig. \ref{spath}. However, in order to include antibound states, 
which is a major aim in this paper, one has to choose
a generalized contour which should enclose not only the Gamow resonances
but also the antibound states. Since for these states it is 
$k_{n}=-i \mid k_n \mid$ the corresponding energy is real and negative.
As we will see, in some circumstances the antibound state has properties
similar to the bound state. However, these states are fundamentally 
different. While the bound state wavefunction diminishes 
exponentially at large distances, the outgoing antibound state diverges 
exponentially.

The antibound states and resonances with
$|\gamma_n|>\kappa_n$ were not included in the completeness relations
originally suggested by Berggren\cite{b68}. This was
related to the regularization
procedure used in that early work.
The first attempt to generalize the completeness by including 
antibound states
and all type of decaying resonances was in a pole RPA approximation 
\cite{ve89}
in which the complex rotation in the radial distance was used as a
regularization method. Later Berggren and Lind also discussed these type 
of generalized completeness relations \cite{be93}.

\begin{figure}
\begin{center}
\includegraphics[width=0.5\textwidth]{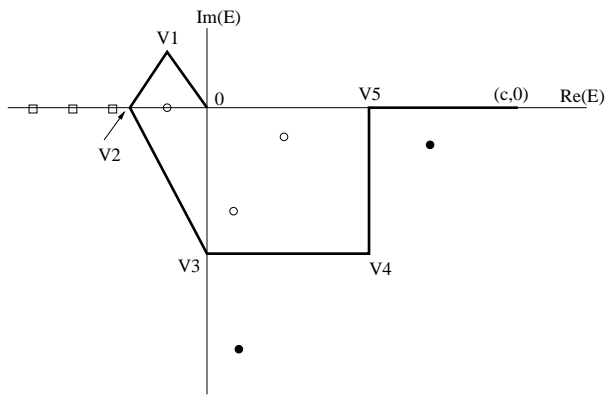}
\vspace{1cm} \caption
{
Generalized integration contour $L^+$ in the complex energy plane
as defined by the vertex points $V_i$. The 
open circles represent the antibound states and the Gamow resonances to 
be included in the Berggren representation, while the solid circles 
are those which are excluded. The open squares are the bound states,
which have to be included independently of the integration contour that
is chosen. The vertex  $(c,0)$ corresponds to
the energy cut-off point $c$. 
}
\label{gpath}
\end{center}
\end{figure}

The single-particle states to be used in the applications
will, therefore, 
be determined by the generalized contour shown in Fig. \ref{gpath}.
The antibound state in this figure is the open circle on the real
energy axis. Notice that the contour lies on the unphysical $E$-sheet since
 $Im(k)<0$.

We will perform the integration of the 
scattering states on the continuum contour by using the Gauss method.
Assuming that there are $N_p$ poles within the chosen contour and $N_g$
Gaussian points in the integration procedure, the set of generalized basis 
states consists of $N=N_p+N_g$ elements. We will order this set such
that $n=1, 2,.., N_p$ labels the poles while $n=N_p+1, N_p+2,..., N$ labels
the sattering states. The corresponding basis vectors are, with standard
notation,
\begin{equation}\label{eq:basis}
\varphi_{nljm}(\vec r)=R_{nlj}(r) (\chi_{1/2}Y_l(\hat r))_{ljm}
\end{equation}
It is important to point out that the Berggren metric affects only the
radial part of these functions while the spin-angular part follows the usual
Hilbert metric.

Summarizing this Section we have presented a representation that defines a
space (Berggren space) which contains as a subspace the Hilbert space.

The formalism dealing with the many-body applications
of the dynamics of a system within the Berggren space is called Complex
Shell Model (CXSM).

The properties of the CXSM are perhaps bizarre and therefore
it is important to
clarify some points that we will use in the Applications. 

The extension of
the Berggren space is defined by the continuum contour. If it is chosen
to be the real energy axis then the resulting CXSM coincides 
with the Shell Model. But as soon as the contour departs from the
real energy axis then a new dimension appears in the Berggren space. However,
whichever is the contour, the Shell Model remains a subspace of the CXSM. 
As a result, all the physical properties evaluated by the standard
Shell Model has to coincide 
with the corresponding quantities evaluated within the CXSM.
In particular, quantities like
transition matrix elements or the angular momentum content of
a state corresponding to bound states should be independent of the contour.
For a complex state this may not be valid any more, since the state may be
outside the Berggren space defined by the chosen contour. For instance 
(and perhaps obvious) the evaluation of those quantities in complex states
by using the real energy axis as a contour would not be possible since the
complex vector is outside the space spanned by the real energy-axis basis 
states (i. e. it is outside the Hilbert space).

Concepts like probabilities have no physical meaning for states outside the
real energy axis, since they become usually complex quantities. However if
the outgoing wavefunction corresponding to the complex poles show 
localization features then one may be able to consider the state as a 
resonance, for which the halflive can be defined.  

These features will be discussed below. 

\section{Applications}
\label{sec:appl}
In this Section we will apply the formalism presented above to weakly
bound nuclei. The most prominent of these nuclei is $^{11}$Li, which
we will also treat here but only as an example of the power of the 
method. Besides, we will show that in the neutron drip-line nucleus 
$^{70}$Ca antibound states play a fundamental role to build the low
energy spectrum.

In all cases the poles as well as the scattering states will be 
evaluated by using the high precision piecewise perturbation 
method \cite{tv2}.

\subsection{The nucleus $^{11}$Li}
\label{sec:li11}
This nucleus has been a playing ground for methods and models
invented to describe features associated to weakly bound nuclei,
specially halos \cite{zhu,gar,suz}. In the shell model approach
one assumes that the odd proton occupying the shell $0p_{3/2}$
is inert and the low lying states can be described as two neutrons
moving outside the core $^{9}$Li. Therefore the ground state of
$^{11}$Li is in this formalism the $0^+_1$ state coupled to the proton
state $0p_{3/2}$, which is a pure spectator providing the angular
momentum of the even-odd nucleus but which otherwise does not 
contribute in any way to the dynamics leading to the low lying states.
The validity of
this approach is justified by the strong correlations between the
valence neutrons at large distance which determine the halo
structure of the nucleus as well as the low energy spectrum \cite{be}.
Moreover, the disregard of the center of mass motion of the core, as we 
will do here, is an approximation which was found to work quite well 
\cite{ebh}.

The core therefore is the nucleus $^9$Li, corresponding to N=6 neutrons, 
with the shells $0s_{1/2}$ and $0p_{3/2}$ frozen while the 
valence shells would be $0p_{1/2}$, $1s_{1/2}$ and perhaps even 
$0d_{5/2}$ and $0d_{3/2}$.
The central field determining these single-particle states is often
chosen as a Woods-Saxon potential. However, in order to reproduce
the amount of s-wave content of the ground state wavefunction in
$^{11}$Li the depth of the potential was taken to be different for
positive- and negative-parity states \cite{ant,be,ebh}. Although
with this choice one gets at the same time large bindings for the 
states $0s_{1/2}$ and $0p_{3/2}$ and weakly bound valence shells,
it is an unpleasant feature to have to use a central potential which is
state dependent. The valence shells are more sensitive to the value
of the central potential close to the nuclear surface, indicating
that a similar feature would be obtained if one uses a standard
Woods-Saxon potential but with an additional central part of
short range and large depth. This would insure a large binding for
the states in the core while the valence shells would be loosely
bound. With this in mind we chose a Woods-Saxon plus Gaussian
central potential. The Woods-Saxon potential is defined by the 
parameters $V_0$=39.97 MeV, $V_0^{so}$=19.43 MeV, 
$r_0=r_0^{so}$=1.27 fm, $a=a^{so}$=0.67 fm, while the Gaussian potential 
is $V(Gauss)=-V_g \exp(-(r/a_g)^2)$ with $V_g$=663 MeV and $a_g$=0.26 fm.
The resulting single-particle states are given in Table \ref{sppot}.

\begin{table}
\caption{
Single-particle states used in the calculation of the two-neutron states 
in $^{11}$Li. The energy $E_n$ and the wave number $k_n$ are related as 
in Eq \ref{eq:cxen}. The $k_n-$value corresponding to the state  
$1s_{1/2}$ shows that this is an antibound state. 
\label{sppot}}
\begin{tabular}{|c|c|c|}
\hline
State & $E_n$ (MeV) &  $k_n$ (fm$^{-1}$) \cr
\hline
$0s_{1/2}$ & (-20.61,0) & (0,0.945) \cr
$0p_{3/2}$ & (-4.525,0) & (0,0.443) \cr
$1s_{1/2}$ & (-0.025,0) & (0,-0.033) \cr
$0p_{1/2}$ & (0.240,-0.064) & (0.103,-0.013) \cr
$0d_{5/2}$ & (4.334,-1.638) & (0.441,-0.081) \cr
$0d_{3/2}$ & (6.396,-9.898) & (0.628,-0.342) \cr
\hline
\end{tabular}
\end{table}

One indeed sees that the shells $0s_{1/2}$ and $0p_{3/2}$ are deeply
bound while the shell $1s_{1/2}$ lies close to threshold, but as an
antibound state, as required by experimental evidence \cite{tz}, and the 
shell $0p_{1/2}$ appears as a resonance at 240 keV and a width 
of 128 keV, as also required by experiment \cite{Boh97}.
These two unbound states are the valence shells. They will determine
the bound ground state of $^{11}$Li, given to the three-body system
consisting of the core, i. e. $^{9}$Li, and the two neutrons its
Borromean character \cite{zhu}.

\begin{figure}
\begin{center}
\includegraphics[width=0.5\textwidth,angle=270]{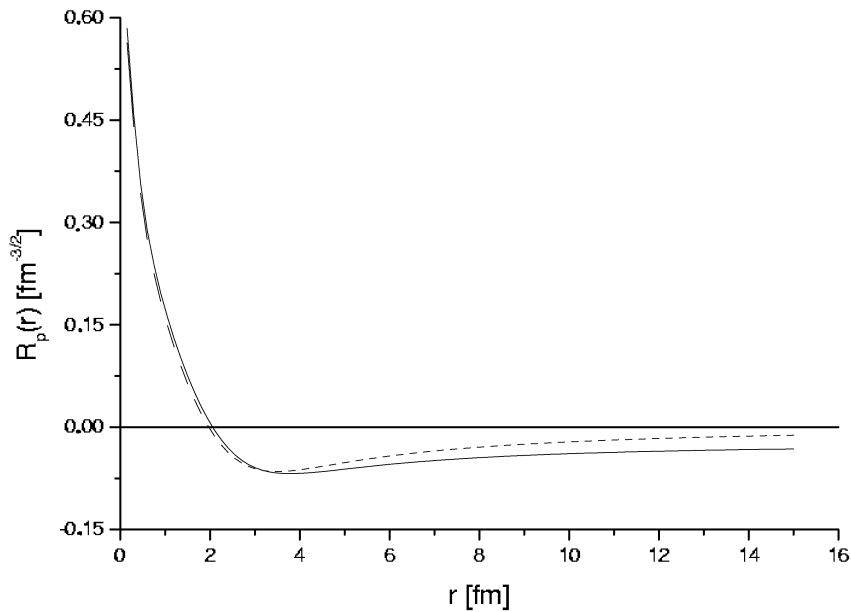}
\vspace{1cm} \caption
{
Radial wavefunction $R_{1s_{1/2}}(r)$ (Eq. (\ref{eq:basis})) 
corresponding to the bound (dashed line) and antibound (straight line)
states in $^{10}$Li at (-0.025,0) MeV. The bound 
state was obtained by changing the depth of the Woods-Saxon potential, as
explained in the text.
The bound (antibound) wavefunction is purely real (imaginary).
}
\label{wfg1s1}
\end{center}
\end{figure}

We are assigning the features of a bound state to the shell $1s_{1/2}$
by labelling it with the principal quantum number $n=1$. This indicates
that it has only one node (excluding the origin)
although it is an antibound state. To show that this is indeed the case
{\it inside} the nuclear volume, i. e. that this complex state is 
localized, we have plotted the radial part $R_p(r)$ 
(with $p\equiv 1s_{1/2}$) of the corresponding
wavefunction in Fig. \ref{wfg1s1}. In this Figure we also plotted
the wavefunction corresponding to the equivalent bound state, i. e.
with the same negative energy.
To obtain the same energy as before but now bound (instead of antibound)
we changed the depth of the Woods-Saxon potential to the value 
$V_0$=42.97 MeV (before it was $V_0$=39.97 MeV). The rest of the 
parameters defining the mean field 
(including the Gaussian part) are the same as before. 

With the usual definition of the nuclear radius, i. e. $R=r_0 A^{1/3}$,
$r_0$=1.25 fm, $A$=10,
one gets $R$=2.69 fm, but in this neutron $l=0$ case
there is not any barrier and the states 
lie so near the continuum threshold that the bound wavefunction extends
far beyond the nuclear radius. This feature motivated the assumption
of a weakly $1s_{1/2}$ bound state in $^{10}$Li when the halo structure
of $^{11}$Li was discovered. However, although great experimental efforts
were made searching for such a state, no trace of it was ever found 
\cite{zhu}. Finally, the $n+^9Li$ system was measured to have
a large and negative scattering length (-17 fm) which prompted
the realization that the state $1s_{1/2}$ indeed exists 
here at low energy,
but actually as an antibound (or virtual) state \cite{tz}. 

In fact there is a similarity between bound and antibound states
lying close to the continuum threshold.
One sees in Fig. \ref{wfg1s1} that for these states 
the wavefunctions
inside the nucleus are practically identical. They depart from each other
only at large distances. Here the antibound wavefunction increases,
eventually diverging. As mentioned above, the CXSM takes care of this
divergence by regularizing all integrals using the complex rotation.

The localization of the antibound state can also be deduced from
the behaviour of the corresponding scattering wavefunction on the
real energy axis.
At the energy $E=\hbar^2 k^2/2\mu$ ($k$ real
and positive) close to a bound or antibound state 
lying at a energy $E_0$ near threshold the scattering 
wavefunction can be written as 
\cite{mig},
\begin{equation}\label{eq:mig}
R_l(kr) \approx \sqrt{\frac{2k|k_0|}{k^2+|k_0|^2}} 
R_l(|k_0|r)
\end{equation}
where $R_l(|k_0|r)$ is the scattering function  at $positive$ energy 
$|E_0|$, although the wave numbers for the bound or antibound state are purely
imaginary, i. e. 
$k_0=\pm i|k_0|$ respectively. That is, the energy values 
for these states are negative, $E_0<0$, but they lie 
on different energy sheets. 

The expression (\ref{eq:mig}) shows that close to threshold 
the radial shape of the scattering wavefunctions is independent 
upon the energy $E$ and the wave function depends on $E$ ($k$) only in its
magnitude through 
the square root factor.
This factor has its maximum at $k=|k_0|$ and therefore the large matrix elements
in an energy region close to $E_0$ are induced by the S-matrix 
pole at imaginary $k_0$. In particular, as seen from
Eq. (\ref{eq:wfa}), the two-body wavefunction in that energy region
will be large on the real energy axis (i. e. within the continuum
shell model). As we will see, this feature explains the large $l=0$
content of the $^{11}$Li(gs) wavefunction. 
The remarkable point in this argument is 
that it does not matter whether the 
pole at $E_0<0$ corresponds to a bound or to an antibound
state. Only the absolute value of $E_0$ is relevant and the 
effect is  the same for both types of states, as expected 
from Fig. \ref{wfg1s1}.

One should not take the similarity between bound and antibound states 
too far. To appraise this we notice that in the presence of 
a high barrier a bound state lying
near the continuum threshold becomes a narrow Gamow resonance if one
changes the central potential adequately, as we did above \cite{Bia01}. 
The imaginary
part of the Gamow wavefunction thus obtained is small while the real
part is very much alike the bound state in a range up to about
twice the nuclear radius. Moreover, the Gamow wavefunctions of all narrow
resonances, not only those lying near the continuum threshold, 
have negligible imaginary parts and real parts which are very similar,
within the nuclear range as above, to the ones provided by a standard 
harmonic oscillator potential \cite{Bia01}. This feature explains why
harmonic oscillator representations have been very successful in
describing observable cluster decays, i. e. very narrow resonances
\cite{Lov98}.
But we want to stress, once again, that in general it is
not the imaginary part of the energy which is a proper 
measure of the width of a resonance but rather the localization of the 
corresponding wavefunction. As we have already discussed, the relation
between $Im(E_n)$ and $\Gamma_n$ is only valid if the width thus
obtained coincides with the Breit-Wigner definition,
i. e. if it satisfies Eq. (\ref{eq:bwiso}). This occurs if the
resonance is isolated, in which case the wavefunction is localized,
as it occurs with narrow Gamow resonances.

\begin{figure}
\begin{center}
\includegraphics[width=0.5\textwidth,angle=270]{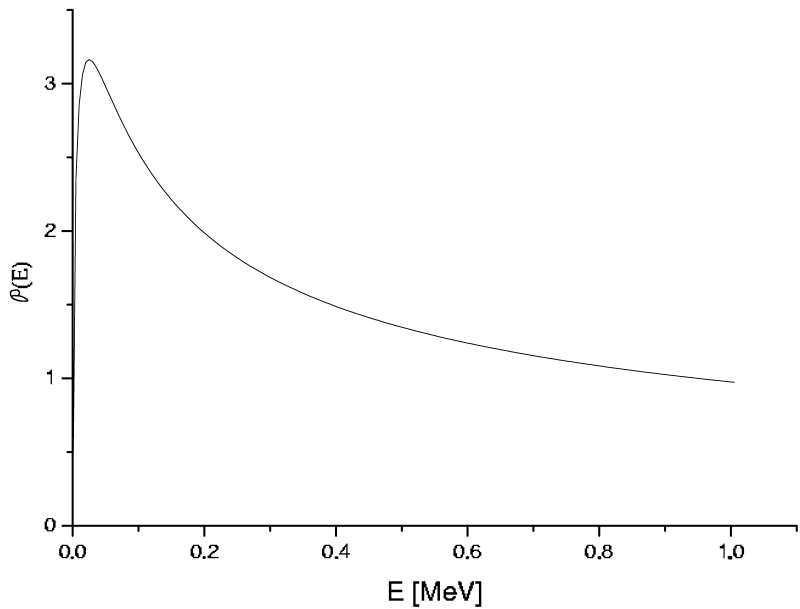}
\vspace{1cm} \caption
{
The function $\frac{{\cal P}(E)}{A}$, Eq. (\ref{eq:prob}), 
corresponding 
to the antibound state at energy $E_0$=-0.025 MeV of Fig. \ref{wfg1s1}.
}
\label{prob}
\end{center}
\end{figure}

The limitations of the definition of the width as twice -$Im(E_n)$
can clearly be seen by trying to apply it to our weakly antibound 
$1s_{1/2}$ state, which cannot have any relation 
with physically meaningful Gamow resonances since there is no barrier 
to trap the
system within the nuclear region (yet, there are Gamow-type $s_{1/2}$ 
poles at bizarre energies, like very large values of -$Im(E_n)$).
Even more, no antibound state can be related to Gamow resonances, since
the energy is purely real. That is,
one cannot recognize the physical meaning of the antibound state by
analysing its properties in the complex energy plane, since here
quantum mechanics is not valid. Instead one has to evaluate meaningful
transitions {\it on the real energy axis}. Thus, the probability
${\cal P}(E)$ that the neutron escapes from the nucleus carrying a 
kinetic energy $E$ is proportional to $|R_l(kr)|^2$, which vanishes at
large radius for bound states. Instead, for antibound states one
gets, from Eq. (\ref{eq:mig}), 
\begin{equation}\label{eq:prob}
{\cal P}(E)=A \frac{\sqrt{E}}{E+E_0}
\end{equation}
where $A$ is an energy independent quantity. As seen in Fig. \ref{prob},
this probability looks like the one corresponding to the decay from
a state at {\it positive} energy $|E_0|$. That is, in a decaying process
the antibound state does not behave as a bound state but rather, again,
as a Gamow resonance but having a width unrelated to the energy. 

\begin{figure}
\begin{center}
\includegraphics[width=0.5\textwidth,angle=270]{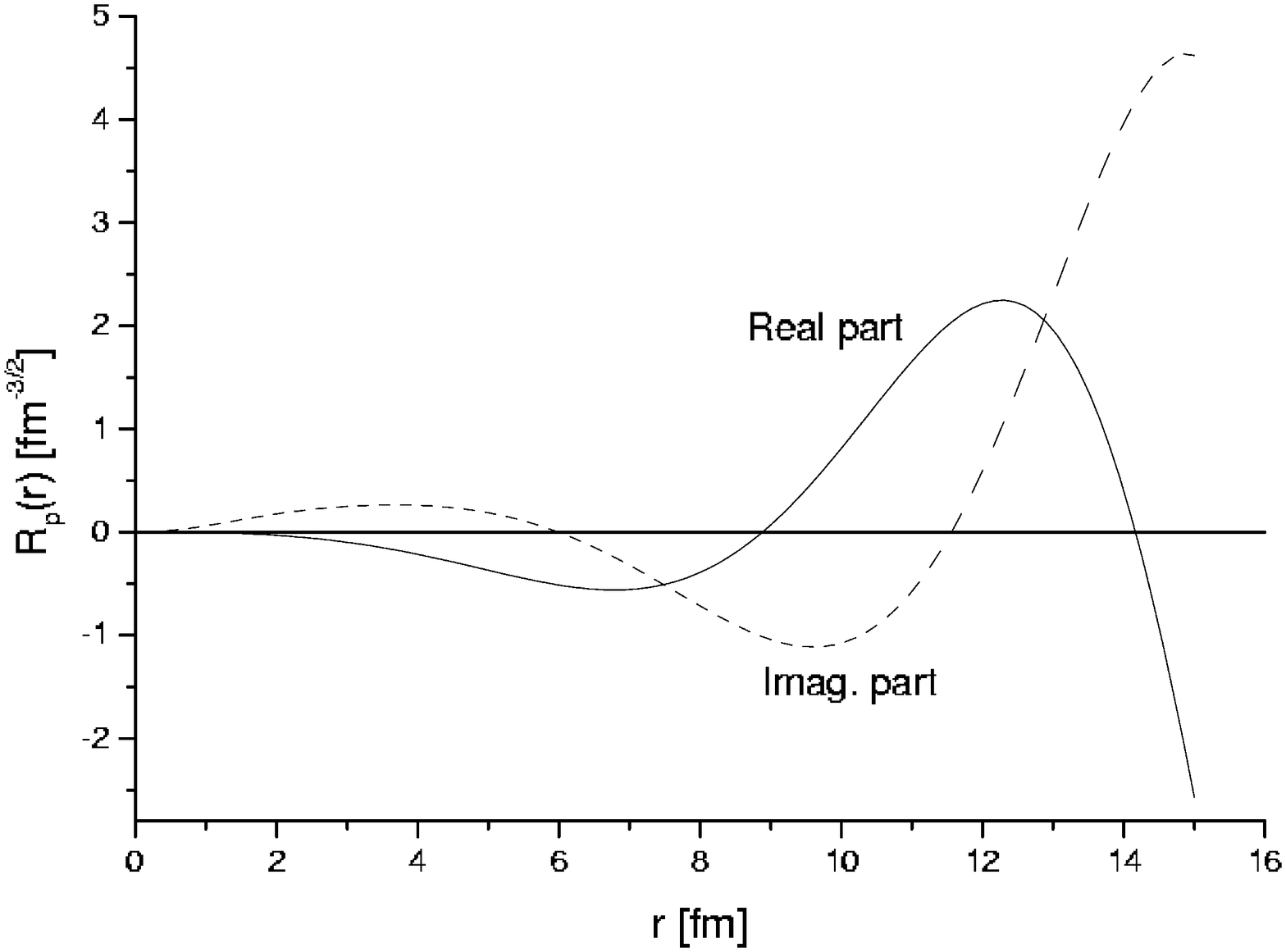}
\vspace{1cm} \caption
{
Radial wavefunction $R_{0d_{3/2}}(r)$ 
corresponding to the "resonance" of Table \ref{sppot}.
Notice the large values of the wavefunction here in comparisson with
those in Fig. \ref{wfg1s1}.
}
\label{wfg0d3}
\end{center}
\end{figure}

The conclusion of this discussion is that poles may have physical
meaning {\it only} if the corresponding wavefunction is localized.
As an illustration of a non-localized pole
we show in Fig. \ref{wfg0d3} the wavefunction corresponding to the
"resonance" $0d_{3/2}$ of Table \ref{sppot}. This state is so wide that
it may be considered part of the continuum background since 
in a region surrounding the nuclear core its value is very small. As 
for the antibound state
discussed previously the wavefunction increases with distance, but the
difference with that case is that now the increase is huge even at a
rather short distance. This state cannot be expected to have any physical
significance. Yet we will include it in the generalized Berggren 
basis. In fact, one of the reasons why we apply in our
calculations high presicion methods \cite{tv2} is to avoid numerical 
errors which would otherwise appear when states with large imaginary 
parts have to be considered.

\begin{figure}
\begin{center}
\includegraphics[width=0.5\textwidth,angle=270]{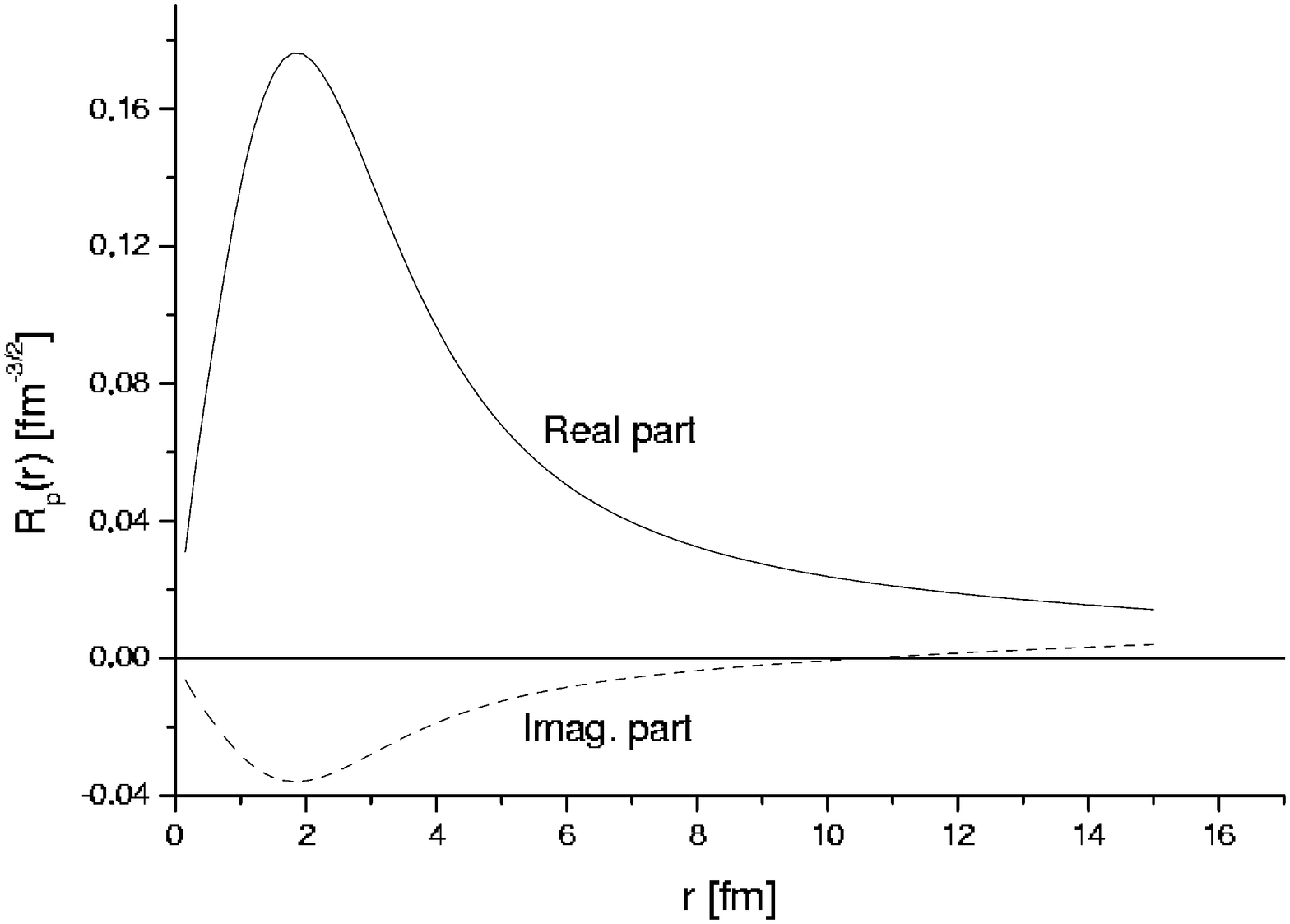}
\vspace{1cm} \caption
{
Radial wavefunction $R_{0p_{1/2}}(r)$ 
corresponding to the single-particle state in Table \ref{sppot}.
}
\label{wfg0p1}
\end{center}
\end{figure}

A state that will be fundamental to build
the low lying spectrum of $^{11}$Li is the Gamow pole $0p_{1/2}$
and, therefore, we show the corresponding wavefunction 
in Fig. \ref{wfg0p1}.
The imaginary part of the energy is in this case not very small
in comparisson to the corresponding real part.
Yet the wavefunction shows clear localization features. This state 
can be considered a resonance.   

One does not need to use the same integration contour for all the
single-particle states, since within the Berggren metric basis states
are independent of each other. It is therefore convenient to use in each
case a contour adapted to the position of the poles. In our case
only for the antibound state it is necessary to have a contour that
lies outside the fourth quadrant in the complex energy plane, as the
one in Fig. \ref{gpath}. For the other partial waves corresponding
to the poles of Table \ref{sppot}
we will use a contour as the one in Fig. \ref{spath}. Moreover, in
each segment $V_i-V_{i-1}$ (with $i>0$ and $V_0=(0,0)$) we choose 
different 
number of points according to how near a pole is to the segment.
The values of the vertices $V_i$ and the number of points $N_i$ between
adjacent vertices defining our Berggren basis is given in Table 
\ref{vert}. The value of $N_i$ is the smallest number of Gaussian
points which provides a convergence of the results of at least 5 digits
in the energies and at least 4 digits in the wavefunctions.

\begin{table}
\caption{
Vertices $V_i$ (in MeV) in the integration contours of
Fig. \ref{spath} corresponding to all partial waves with values
of $(l,j)$ as those in the states of Table \ref{sppot} except 
for the partial wave $s_{1/2}$ for which the contour shown in Fig. 
\ref{gpath} was used. $N_i$ is the number of Gaussian integration 
points in the segment between the vertices $V_i$ and $V_{i-1}$. 
The point $V_0$ (not shown in the Table neither in the Figures) is 
the origin, i. e. $V_0=(0,0)$. The energy cut-off point is 
$(c,0)$ = 10 MeV and the number of Gaussian points between the last 
vertex and the cutt-off point $(c,0)$ is 4 in all cases.
\label{vert}}
\begin{tabular}{|c|c|c|c|c|c|c|c|c|c|c|}
\hline
partial wave & $V_1$ & $N_1$ & $V_2$ & $N_2$ & $V_3$ & $N_3$ & 
$V_4$ & $N_4$ & $V_5$ & $N_5$ \cr
\hline
$s_{1/2}$ & (-0.025,0.1) & 10 & (-0.050,0) & 10 & (0,-0.7) & 
20 & (3,-0.7) & 6 & (3,0) & 2 \cr
$p_{3/2}$ & (0,0) & 2 & (0,-0.7) & 4 & (3,-0.7) & 4 & (3,0) & 4
& ------ &---\cr
$p_{1/2}$ & (0,0) & 2 & (0,-0.7) & 6 & (3,-0.7) & 10 & (3,0) & 6 
& ------ &---\cr
$d_{5/2}$ & (3.5,0)& 4 & (3.5,-3) & 4 & (5,-3) & 4 & (5,0) & 4
& ------ &---\cr
$d_{3/2}$ & (5,0) & 4 & (5,-10.7) & 4 & (8,-10.7) & 4 & (8,0) & 4 
& ------ &---\cr
\hline
\end{tabular}
\end{table}

The valence poles plus the scattering states thus defined constitute
our generalized single-particle Berggren basis. 

The two-particle basis is built from the single-particle states as usually.
To evaluate the two-body energies and corresponding wavefunctions one has
to solve the dispersion relation (\ref{eq:disrel}). To determine the
strength $G_\alpha$ we will use the standard procedure of adjusting its
value to fit the lowest energy of the two-particle state carrying angular
momentum $\lambda_\alpha$ (and parity $(-1)^{\lambda_\alpha}$, since the 
force is separable ).

The only two-neutron positive-parity state for which there is any experimental 
data in $^{11}$Li is the ground state, i. e. $\lambda_\alpha$=0 (although  
due to the presence of the 
proton the real spin of $^{11}$Li(g.s.) is $3/2^-$). The corresponding 
experimental
energy is $\omega_{n_\alpha=1,\lambda_\alpha=0}\approx$-0.295 MeV \cite{gar}.

The short range and attractive character of the residual nuclear interaction
causes the lowest $0^+$ state to lie below the 
lowest two-particle configuration. Therefore the vertex $V_2=(a,0)$ in 
Fig. \ref{gpath} has to fulfill the condition $\omega_{0^+_1}<2a$. 
This restricts the position of the vertex $V_2$ to the range
$-0.1 MeV < a < -0.025 MeV$. As seen in Table \ref{vert} we use  
$V_2=(-0.050,0)$ MeV which is very close to the energy of the antibound
state. This is not an optimal situation since close to a pole the scattering
waves increase rapidly as the energy approaches the pole. In fact, one of 
the advantages of using the representation (\ref{eq:delb}) for the Dirac
delta function is to avoid the numerical difficulties that one encounters
when integrating on the real energy axis in the presence of very narrow 
resonances. These difficulties can be overcome by choosing an 
integration countour 
lying far enough from the resonance energy \cite{shec}. In our case the
contour has to be near the pole and therefore we had to use many integrations 
points around the antibound state. This explains the relatively 
large values of $N_1$, $N_2$ and $N_3$ corresponding to the partial wave 
$s_{1/2}$ in Table \ref{vert}.
 
Another undesirable feature of this situation is that the two-particle
configurations corresponding to the scattering states may acquire values 
which are very large (according to Eq. (\ref{eq:wfa})) and therefore 
unfamiliar. But this is not a problem in 
itself, since the wavefunction components
in the complex energy plane depends upon
the contour that one uses. It is worthwhile to point out again that the 
results have physical meaning only when evaluating quantities defined 
on the real energy axis, as for instance the energies of bound states 
or the angular momentum content of the two-particle wavefunctions of 
those bound states. The ground state of $^{11}$Li lies at about 
-295 keV, as mentioned above, and the angular momentum content is about 
60 \% $s$-states and 40 \% $p$-states, although some small components
of other angular momenta are not excluded \cite{gar}. 

Notice that for bound states 
the angular momentum content $A_l$ does not depend upon the
contour because it is a sum over all basis states. Thus, for states with
angular momentum $\lambda_\alpha$ = 0 it is 
\begin{equation}\label{eq:amc}
A_l= \sum_{nn'l'\not=l j}X(nl'j,n'l'j;\alpha)^2 
\end{equation}
which is a real number if the state $\alpha$ is bound 
and complex otherwise since the real energy axis is excluded 
as a contour to describe complex states. This also implies that 
$A_l$ (as well as all physical quantities corresponding to "resonances"
which are not isolated) may depend upon the contour if this contour
defines a Berggren space which does not fully contain the "resonance".
However, due to the normalization of the state it is
$\sum_l A_l=1$ in all cases.

The imaginary part of $A_l$ may give a measure of how significant (isolated) 
is the complex state. If $|Im(A_l)|$ is large then the state may 
be considered a part of the two-particle continuum background. Again here,
it is not very clear what is meant by "large" and it would be more convenient
to answer this question by looking at the localization of the corresponding
wavefunction, as we will do below.

The property that $A_l$ for bound states do not depend upon 
the contour has been used by us as an important test for checking 
our computer codes. 

The energy of $^{11}$Li(gs) will be fitted to
adjust the strength $G_\alpha$ of the separable interaction. Therefore
it is $G_\alpha$ which should remain the same and a real number
(since the Hamiltonian is Hermitian) as one changes the contour.

Using for the field (\ref{eq:field}) the parameters $f_0$=10,
$R'$=4.5 fm and $a'$=1.5 fm we evaluated within the generalized 
Berggren basis defined above the strength from Eq. 
(\ref{eq:disrel}) to obtain the value $G_0$=0.1659 MeV. Using
as a contour the real energy axis we obtained the same value, 
confirming the formalism as well as the precision of our computer
codes.

The dispersion relation has as many solutions as the dimension of the  
two-particle basis. The overwhelming majority of these solutions
form part of the continuum background. These continuum states are easy to 
recognize because they feel the
interaction very weakly and therefore they lie very close to their zeroth
order energy, i. e. they are ordered according to the lines defining
the contour. For illustrations of this feature see Ref. \cite{cxsm}.
The physically meaningful states are localized and, therefore, one
expects that the main wavefunction configurations of the relevant two-particle 
states should correspond to bound states or resonances.
One indeed finds this feature in the main wavefunction components of
the ground state (i. e. $0^+_1$) at -0.295 MeV and in a state at (0.274,-0.247) MeV 
($0^+_2$) which may 
be a resonance. The largest of those components are
$$
|0^+_1>=(1.79,0)|(1s_{1/2})^2>+(0.15,1.02)|(1s_{1/2}c_1)>  
+(0.03,0.98)|(1s_{1/2}c_2)>
$$
$$
|0^+_2>=(0.70,0.12)|(0p_{1/2})^2>+(-0.50,-0.08)|(0p_{1/2}c_3)>  
+(0.36,-0.15)|(0p_{1/2}c_4)>
$$
where $c_i$ are states belonging to the continuum contour, i. e. they
are scattering waves with angular momenta $(l,j)$ and energies 
$\epsilon_i$ (in MeV) given by
$$
|c_1>=|l=0,j=1/2,\epsilon_1=(-0.048,-0.031)>; ~~
|c_2>=|l=0,j=1/2,\epsilon_2=(-0.046,-0.056)>
$$
$$
|c_3>=|l=1,j=1/2,\epsilon_3=(0,-0.266)>; ~~
|c_4>=|l=1,j=1/2,\epsilon_4=(0,-0.119)>
$$
These continuum states are important in the description of our relevant
$0^+$ levels because the zeroth order energies of the corresponding 
two-particle basis states are very close to the correlated energies.
However, the
CXSM wavefunction components depend upon the contour one chooses and, 
therefore, they have physical meaning only if the contour coincides with 
the real energy axis, as in the continuum shell-model. This feature 
is valid even if the state is bound, although in this case the wavefunction 
itself does not depend upon the contour and it has the normal physical 
meaning of quantum mechanics, as we wil see below. 

Since the components of the CXSM wavefunction are not well defined quantities
it may seem unreasonable to probe features like localization by analysing
such components, as we did above. However we have found that physically
relevant states always have wavefunctions with main components consisting of 
bound or resonant single-particle states, independently of the contour one 
chooses. In our case, the quantity which has a physical (i. e. quantum 
mechanical) significance is the angular 
momentum content $A_l$, Eq. (\ref{eq:amc}). We found that for the ground
state $A_l$ is, in  percentage, 49 \% s-states, 47 \% p-states and 4 \% 
d-states, in agreement with experiment. 

Although for bound states $A_l$ is real its components may be complex, contour
dependent and
with very large absolute values. For instance, in $^{11}$Li(gs) the 
contribution of the 
pole-pole component $|(1s_{1/2})^2>$ is (3.218,0.004), while the 
pole-continuum components $|1s_{1/2} c_i>$, where $c_i$ is an s-wave on the 
contour, add up to 
(-8.473,-0.009) and the continuum-continuum components 
$|c_ic_j>$ to (5.746,0.005). The corresponding values for the p-waves is
(0.598,-0.202) for the pole-pole component $|(1p_{1/2})^2>$,
(-0.146,0.218) for the pole-continuum components $|1p_{1/2} c_i>$
and (0.001,-0.016) for the continuum-continuum components. The 
$p_{3/2}$, $d_{3/2}$ and  $d_{5/2}$ contribution
adds up to a total of (0.060,0). 
The total sum are the values 0.49 for the s-waves, 0.47 for the p-waves
and 0.04 for the d-waves, as quoted above. That is, the
total sum is real and represents a probability while the partial
components have no meaning and are contour dependent.
Notice that this is valid even for the pole-pole component, since its value
is proportional to the wavefunction amplitude $X$ which depends upon the
contour through the normalization constant (see Eq. (\ref{eq:wfa})).

In a calculation performed on the real energy axis there is neither pole-pole 
nor pole-continuum contributions (since there is no bound single-particle
state in this case) and all partial contributions 
are continuum-continuum components that are probabilities, i. e. they are 
real, positive and less than or equal to unity. These values coincide with the 
percentages of $A_l$ given above, once again checking the reliability of our 
codes.

For the $0^+_2$ state all partial contributions to $A_l$ are 
complex although they have to add to 
unity due to the normalization of the CXSM wavefunction. One thus obtains 
that $A_l$ is (13,-10) \% s-states and (87,10) \% p-states, indicating that
this "resonance" is mainly built upon the $0p_{1/2}$ single-particle 
resonance. 

It is interesting to see how the mixing among the various components of the
angular momentum content evolves as the strength of the interaction increases
from the zeroth order value where the states are indeed
given by the poles, i. e. for $G=0$ it is 100\% s-states for $0^+_1$ and 
100\% p-states for $0^+_2$. 
In Table \ref{amcg} we present the evolution of these values, as well as the
energies of the states, as a function of the strength $G$.

\begin{table*}
\caption{
Energies E and angular momentum content $A_l$, Eq. (\ref{eq:amc}), for the
angular momenta $l=0, 1$ and $2$ ($s, p$ and $d$), corresponding to the 
states $0^+_1$  and $0^+_2$ in $^{11}$Li as a function of the interaction 
strength $G$. The values of E and G are in Mev while $A_l$ is $\times$ 100.
The strength fitting the available experimental data is $G_0$=0.1659 MeV. 
\label{amcg}}
\begin{ruledtabular}
\begin{tabular}{|c|cccc|cccc|}
&\multicolumn{4}{c}{$0^+_1$}&\multicolumn{4}{c|}{$0^+_2$}\cr
\hline
G & E & s & p  & d  & E  & s & p & d   \cr
\hline
0.01  & -0.061 & 100 & 0 & 0 & (0.466,-0.117) & 0 & (100,0) & 0\cr
0.05  & -0.091 & 100 & 0 & 0 & (0.388,-0.089) & (-4,0) & (104,0) & 0\cr
0.09  & -0.092 & 100 & 0 & 0 & (0.292,-0.144) & (-14,-12) & (114,12) & 0\cr
0.13  & -0.102 &  99 & 1 & 0 & (0.261,-0.224) & (14,-9) & (86,9) & 0\cr
0.17  & -0.328 & 48 & 47 & 5 & (0.271,-0.276) & (17,-4) & (83,4) & 0\cr
\end{tabular}
\end{ruledtabular}
\end{table*}

Looking at the imaginary values of these quantities as well as
at the energy of the predicted $0^+_2$ state (i. e. (0.274,-0.247) MeV), 
which is large, one may doubt whether this state is indeed
a resonance and therefore whether it
could be detected experimentally. To decide upon this we will analyse its
localization properties.
But first we have to learn which are the features that determine such
localization in this weakly bound nucleus. This can be done by studing 
the two-particle wavefunction of the bound ground state.
This would help us to decide whether the state $0^+_2$ is indeed a resonance
by comparing the properties of its wavefunction with those of the ground state.
A convenient way of doing this is by exploiting the clustering features of
the ground state wavefunction in the singlet (S=0) component \cite{jan83}.
We will call $S(r)$ this singlet part of the wavefunction. Its expression can 
be found in Ref. \cite{jan83}.

Due to the two-neutron clustering one can recognize whether the center of mass 
of the two-neutron system remains inside the nuclear core by analysing 
$S(r)$ in some direction, for instance in the x-direction. That is we will
choose $r=r_1=r_2$, where $\vec r_i=(r_i,\theta_i,\phi_i)$ is the 
coordinate of the particle $i$ with $\theta_i=\pi/2, \phi_i=0$.
Notice that this direction is irrelevant since the system is spherically 
symmetric.

\begin{figure}
\begin{center}
\includegraphics[width=0.5\textwidth,angle=270]{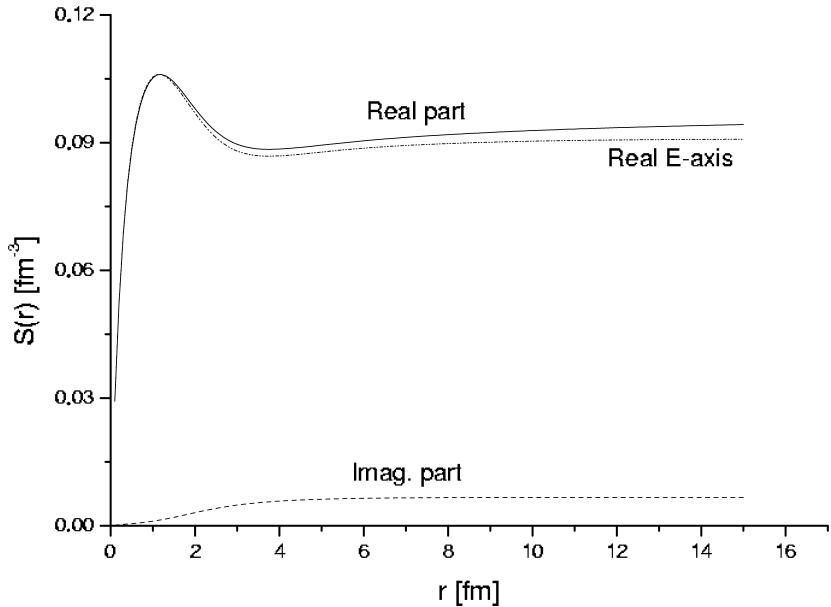}
\vspace{1cm} \caption
{
Two-particle radial wavefunction S(r) corresponding to the singlet component 
of the ground state of $^{11}$Li evaluated by using the generalized Berggren
representation. The dashed line is the same wavefunction but using as
a representation the real energy axis, i. e. calculated
within the Continuum Shell Model framework.
}
\label{tpwf0p1}
\end{center}
\end{figure}

In Fig. \ref{tpwf0p1} we show the function $S(r)$ corresponding to 
the ground state of $^{11}$Li.
One sees that inside the nucleus the wavefunction of the two valence particles 
is highest and the corresponding imaginary part smallest, indicating that 
there is a localization. Since this
is the quantum mechanical wavefunction corresponding to the bound system, its
value should not only be independent of the contour but also it should
represent a probability amplitude. Therefore the small imaginary part that
appears at long distance has to be considered a limitation of the calculation.
One also sees that
this wavefunction extends far outside the nucleus, as expected since this
is the feature that causes the halo. Yet, one may think that the imaginary
part as well as the large value of the real part at large distances is an 
effect of the diverging character of the
complex wavefunctions in the generalized Berggren basis. To be sure that 
this is not the case, and considering that this wavefunction has a physical
meaning, we calculated it again but using the real energy axis as a 
representation. As seen in the Figure, we found that the wavefunction
thus calculated is virtually the same as the one using the generalized
Berggren representation. As with the imaginary part, there are small
differences at large distances, and this has to be attributed to
computational limitations that makes it difficult to treat exactly
the diverging character of the single-particle states.

As a final comment, it has to be noted that although the two-particle 
wavefunction  extends far in space (a feature which is known, see e. g. 
Fig. 1 of Ref. \cite{be}), the mean square radius of the nucleus
is not excessively large since this quantity is strongly influenced by
the particles in the core, including the three protons \cite{be}. The
extension of the wavefunction corresponding to the valence particles
provides a tiny contribution to the total wavefunction of the nucleus,
which is the reason why the density of the halo is low. 

\begin{figure}
\begin{center}
\includegraphics[width=0.5\textwidth,angle=270]{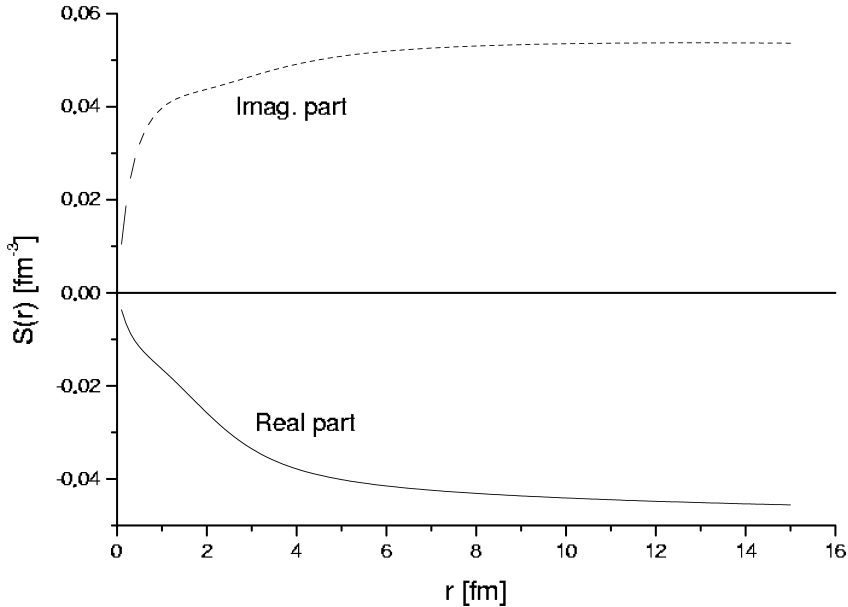}
\vspace{1cm} \caption
{
Two-particle radial wavefunction S(r) corresponding to the singlet component 
of the state $^{11}$Li($0^+_2$) evaluated by using the generalized Berggren
representation. 
}
\label{tpwf0p2}
\end{center}
\end{figure}

We are now in a position to analyse the wavefunction of the $0^+_2$ state.
We thus plotted in Fig. \ref{tpwf0p2} the singlet function S(r) for this
state. We see that now the wavefunction
does not show any localization feature inside the nucleus. Perhaps even
worse, the imaginary part is very large, thus providing a large imaginary
part of the "probability" if it would be assigned a quantum mechanical
meaning.
This state is not a resonance and therefore it is not surprising that 
it has not been observed.

\begin{figure}
\begin{center}
\includegraphics[width=0.5\textwidth,angle=270]{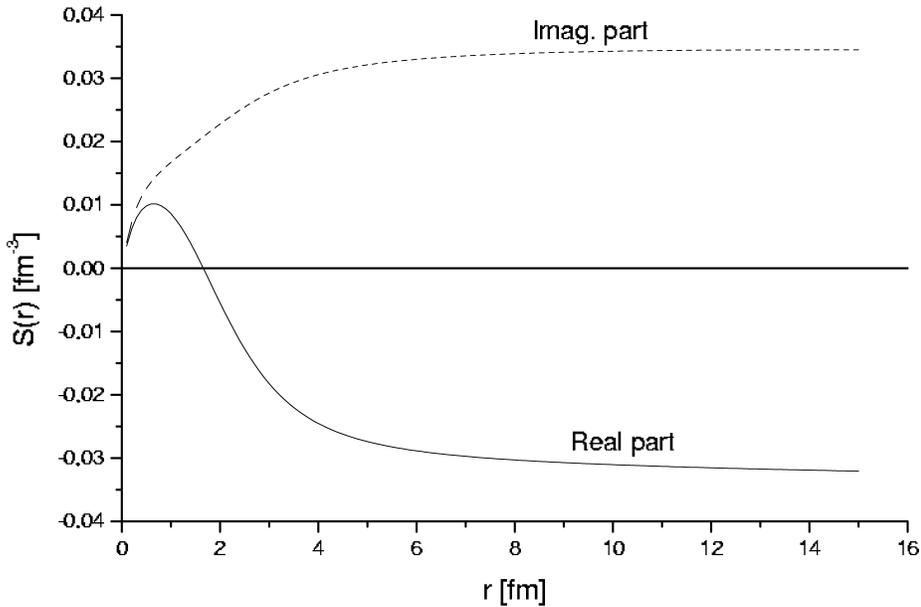}
\vspace{1cm} \caption
{
Two-particle radial wavefunction S(r) corresponding to the singlet component 
of the state $^{11}$Li($0^+_2$) evaluated by using the generalized Berggren
representation with a G=0.05 MeV interaction. 
}
\label{tpwf0p2g05}
\end{center}
\end{figure}

The narrowest $0^+_2$ state in Table \ref{amcg} corresponds to G=0.05 MeV. 
One may
thus learn from this case whether its small imaginary part of the energy
implies that it is localized. 
We thus plotted in Fig. \ref{tpwf0p2g05} the corresponding function S(r).
One sees that something very strange happens. While the real part of the
wavefunction shows localization the imaginary part does not. Again, the
criterion of defining a resonance according to its localization features 
shows that neither this is a physically meaningful state.
Yet, it is remarkable that as $G$ increases this state 
first becomes narrower and
then, just after $G$=0.05 MeV in the Table, its width increases rather
fast. This indicates that at this value of $G$ there is a mechanism
that moves the state down in the complex energy plane. To examine this
feature we show in Table \ref{amcgsp} the dependence of the angular
momentum content for each partial wave upon $G$. We show only the 
$s-$  and $p-$components since the other angular momenta play no role
here. One sees in the Table that the mixing between these components is
not as strong as the mixing between the pole-pole $|(1p_{1/2})^2>$
component and the pole-continuum components. It is also conspicuous
that the continuum-continuum $l=1$ partial waves has virtually no
influence. One therefore can conclude that the reason why the state
does not become narrower as the interaction increases is that the
coupling of the single-particle resonance to the continuum becomes
stronger with an stronger interaction and, as a result, the two-particle 
state is itself pushed towards the continuum. At no value of $G$ the
two-particle system in the $0^+_2$ complex state is trapped by the 
interaction inside the nucleus and the state never becomes a resonance.

\begin{table*}
\caption{
Pole-pole (PP), pole-continuum (PC)
and continuum-continuum (CC) contributions to
the $s$ and $p$ angular momentum content $A_l$
corresponding to the state $0^+_2$ in $^{11}$Li 
as a function of the interaction strength $G$. The values of G are in Mev 
while $A_l$ is $\times$ 100.
The strength fitting the available experimental data is $G_0$=0.1659 MeV. 
\label{amcgsp}}
\begin{ruledtabular}
\begin{tabular}{|c|ccc|ccc|}
&\multicolumn{3}{c}{$s-content$}&\multicolumn{3}{c|}{$p-content$}\cr
\hline
G & PP & PC  & CC  & PP & PC & CC   \cr
\hline
0.01  & 0 & 0 & 0 & (100,0) & 0 & 0\cr
0.05  & (2,0) & (-4,5) & (-2,-5) & (104,2) & (0,-2) & 0\cr
0.09  & (3,14) & (-31,-4) & (14,-22) &  (101,20) & (13,-8) & 0\cr
0.13  & (-8,-3) & (8,-17) & (14,11) &  (36,30) & (50,-21) & 0\cr
0.17  & (-6,-4) & (12,-16) & (11,16) &  (39,20) & (44,-16) & 0\cr
\end{tabular}
\end{ruledtabular}
\end{table*}

Finally, it is interesting to analyse a $1^-$ state which
was reported to be found
at about 1.3 MeV and a width of about 0.75 MeV in a $^{11}Li + p$ 
experiment \cite{kor97}. From Table \ref{sppot} one sees that
this state lies too low to be a particle-hole excitation, since
the lowest configuration of this type is $|1s_{1/2}0p_{3/2}^{-1}>$
at an energy of 4.5 MeV, and it lies too high to be a particle-particle
state, since the lowest two-particle configuration is
$|1s_{1/2}0p_{1/2}>$ at an energy of 0.215 MeV. Therefore we did not find
any reasonable G value, even assuming a repulsive force (i. e. G real and
negative) that could provide this state within our shell-model approach.

\subsection{The nucleus $^{70}$Ca}
\label{sec:ca70}
\begin{figure}
\begin{center}
\includegraphics[width=0.5\textwidth,angle=270]{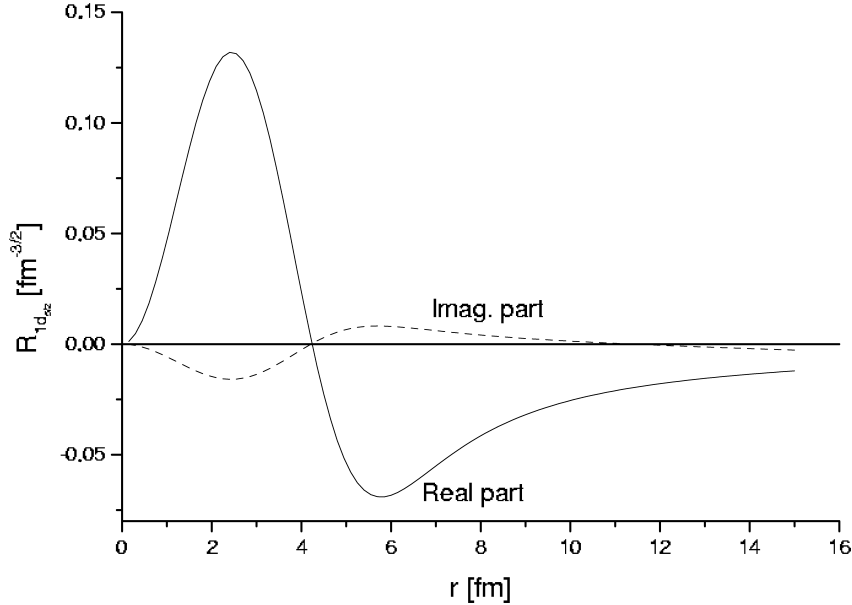}
\vspace{1cm} \caption
{
Radial wavefunction $R_{1d_{5/2}}(r)$ 
corresponding to the single-particle state in Table \ref{speca}.
}
\label{wfg1d5ca}
\end{center}
\end{figure}

We have seen in the previous Subsection that in $^{11}$Li(gs) the halo
is induced by the extension of the wavefunctions corresponding to the
valence single-particle states. This, in turn, is due to the lack of 
a barrier that would hold the particles tightly bound to the core. It
is a delicate balance, since without the pairing interaction 
acting upon the valence particles the nucleus would not be bound, i. e.
the ground state wavefunction would not be localized. One may therefore 
wonder whether valence single-particle states larger than $l=1$
would still induce halos. With this aim in mind we tried to find such
a case by following the trend of single-particle states in a
relativistic mean field calculation. We thus found that the nucleus
$^{72}$Ca (corresponding to the core 
Z=20, N=50) fulfills the conditions that we look for,
since the valence single-particle states are again an antibound $2s_{1/2}$
shell but now the next shell is $2d_{5/2}$. In order to simulate the order
of the single particle states given by the relativistic calculations
we used a Wood-Saxon potential defined by
$a$ = 0.67 fm, $r_0$ = 1.27 fm,  $V_0$ = 39 MeV and $V_{so}$ = 22 MeV.
The corresponding single-particle states are shown in Table
\ref{speca}. 
One sees that the antibound state as well as the first excited state lie
near the energy threshold, as in the previous case. Even the wavefunction
of the antibound state is similar to the one in $^{10}$Li, as expected.
However, since now the first excited state carries a higher angular momentum 
as compared with the previous case, one would expect that the corresponding
wavefunction would be too much localized, thus hindering the formation of the
halo. As seen in Fig. \ref{wfg1d5ca} this is not the case since the 
wavefunction extends much beyond the standard value of the nuclear radius 
in $^{71}$Ca, i. e. $R$=5.18 fm.

\begin{table}
\caption{
Valence single-particle states used in the calculation of the 
two-neutron states 
in $^{72}$Ca. The energy $E_n$ and the wave number $k_n$ are related as 
in Eq \ref{eq:cxen}. The $k_n-$value corresponding to the state  
$2s_{1/2}$ shows that this is an antibound state. 
\label{speca}}
\begin{tabular}{|c|c|c|}
\hline
State & $E_n$ (MeV) &  $k_n$ (fm$^{-1}$) \cr
\hline
$2s_{1/2}$ & (-0.056,0) & (0,-0.051) \cr
$1d_{5/2}$ & (0.488,-0.053) & (0.153,-0.008) \cr
$1d_{3/2}$ & (2.089,-1.545) & (0.334,-0.110) \cr
$0g_{7/2}$ & (6.772,-0.748) & (0.568,-0.031) \cr
$0h_{11/2}$ & (5.386,-0.106) & (0.506,-0.005) \cr
\hline
\end{tabular}
\end{table}

There is another important difference between the single-particle states
in Tables \ref{sppot} and \ref{speca}, namely the appearance of
the narrow high spin resonance $0h_{11/2}$. This may induce a high-lying
localized two-particle resonance.

To probe the differences, if any, between $^{11}$Li and $^{72}$Ca we proceeded
as in the previous subsection and solved the generalized CXSM in this case
by using the separable interaction provided by the field $f$ of Ref. 
\cite{ant}. As already reported in that reference, we thus found that the 
ground state two-particle wavefunction has the same features as the 
corresponding one in $^{11}$Li. 

\begin{figure}
\begin{center}
\includegraphics[width=0.5\textwidth,angle=270]{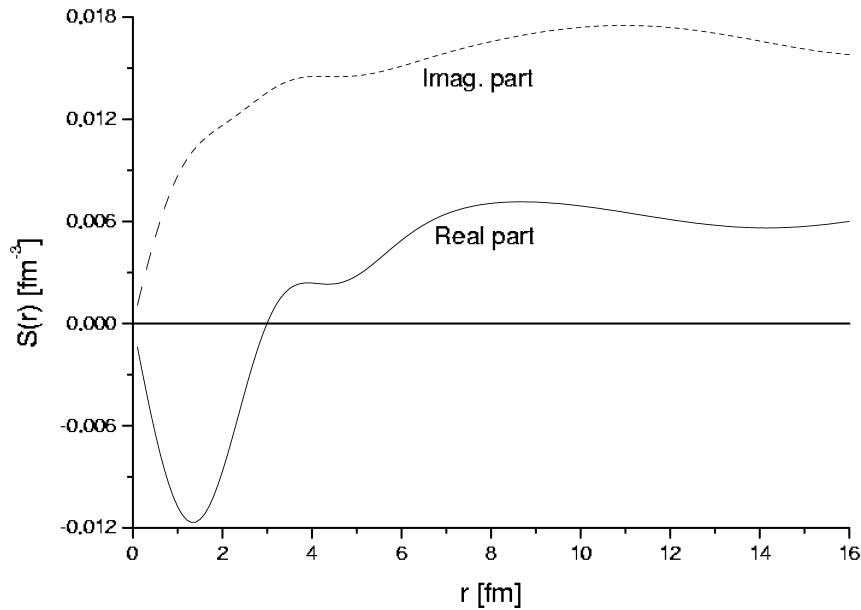}
\vspace{1cm} \caption
{
Two-particle radial wavefunction S(r) corresponding to the singlet component 
of the state $^{72}$Ca($0^+_2$) evaluated by using the generalized Berggren
representation. 
}
\label{tpwf0p2ca}
\end{center}
\end{figure}

There is also a second $0^+$ state lying at (0.550,-0.350) MeV. which
in Ref. \cite{ant} was considered to be a likely resonance. We have now
the possibility to examine whether this is the case by looking at the 
wavefunction $S(r)$. The notable peculiarity of this wavefunction is that
it looks very similar to the corresponding one in $^{11}$Li, as can be seen
by comparing Figs. \ref{tpwf0p2} and \ref{tpwf0p2ca}.
It thus seems that the nuclei $^{11}$Li and $^{72}$Ca are both halo nuclei
although the valence wavefunctions correspond to different number of nodes
and even orbital angular momenta. 
This shows, once more, that the characteristic determining the formation of 
halos is the extension of the valence wavefunctions in space, and this may 
happens even if the valence particles move in high angular momentum orbits. 

We have also investigated whether it appears a narrow two-particle resonance
as a result of the coupling of the particles in the single-particle state
$0h_{11/2}$. This we did not observe by using the strength $G_0$ that provides
the bound ground state of $^{72}$Ca. We then decided to follow the trajectory
in the complex energy plane of the state $|0h^2_{11/2}>$ as $G$ increases 
from its zeroth order value. As in Ref. \cite{cxsm} we found that this state
indeed becomes narrower as $G$ increases to a certain point, but afterwards
the coupling of one of the particles to the continuum increases the width
of the resonance which  eventually becomes itself part of the continuum
background. We have already seen this feature above when analysing the state
$0^+_2$ in $^{11}$Li and it appears again in the halo nucleus $^{72}$Ca
as well as in the drip line nuclei analysed previously by us \cite{cxsm}.
This teory thus predicts that only if the pairing interation is relatively
weak two-particle resonances would appear as the result of the coupling
of the particles to high lying narrow single-particle resonances.

\section{Summary and conclusions}
\label{sec:sum}
In this paper we have presented a new formalism to evaluate
microscopically two-particle resonances in halo nuclei by 
using a generalization of the complex shell model (CXSM) 
presented in Ref. \cite{cxsm}.
Since one of the elements inducing the formation of halos was found to be
the appearance of antibound states lying close to the continuum threshold,
this generalization consists in defining a contour in the 
complex energy plane which comprises the antibound states.
That is, the antibound states
and the Gamow states (i. e. states lying in the complex energy plane)
are selected by appropiate contours
in the complex energy plane. 
Terefore within this formalism bound states, antibound states,
Gamow states  and the
continuum background are all basis single-particle states treated on the 
same footing.
These states are poles of the corresponding single-particle Green function.

The contribution of the pole-pole, pole-continuum and continuum-continuum 
configurations in the two-particle systems can be easily analysed.
The effects induced by antibound states and the continuum encircling the
poles can be studied separately. 

Given the rather unfamiliar characteristics of both the CXSM and the
antibound states we have described in detail those characteristics.
In particular, we have shown that the effects induced by antibound states 
lying close to threshold upon physically measurable quantities is exactly
the same as those provides by bound states lying also close to threshold.
In both cases the main feature is that the single-particle wavefunction
extends far in space. 

We have also analysed the properties that a many-particle state lying on
the complex energy plane may have to be considered a "resonance", i. e. 
a measurable state appearing in the continuum part of the spectrum.
We have thus found that the resonant wavefunction has to be localized within 
the nuclear system. In order to illustrate these features as well as to show
the advantages of the formalism we evaluated
two-particle states in the well known case of $^{11}$Li as well as in 
the nucleus $^{72}$Ca. We analysed the localization properties of the
two-particle wavefunctions in these nuclei trying to find high lying 
resonances. However we did not find evidences that would indicate that
such resonances exist in these cases.

Finally, we have found that the theory predicts that only if the 
two-body interation is relatively
weak two-particle resonances would appear as the result of the coupling
of the particles to high lying narrow single-particle resonances.

\acknowledgments
This work has been supported by FOMEC and Fundaci\'on
Antorcha (Argentina), by
the Hungarian OTKA fund Nos. T037991 and T046791 and by
the Swedish Foundation for International Cooperation
in Research and Higher Education (STINT).

\end{document}